\documentclass[12pt]{amsart}

\title{Homeostasis in Chemical Reaction Pathways}

\author{V. Malyshev, A. Manita , A. Zamyatin}

\makeatletter
 \theoremstyle{plain}    
 \newtheorem{thm}{Theorem}[section]
 \numberwithin{equation}{section} 
 \numberwithin{figure}{section} 
 \newtheorem{cor}[thm]{Corollary} 

\usepackage{a4wide}

\usepackage{graphicx}

\usepackage{amsmath}
\usepackage{amsfonts}
\usepackage{amssymb}%
\theoremstyle{plain}
\numberwithin{equation}{section}
\numberwithin{figure}{section}
\theoremstyle{plain}
\numberwithin{equation}{section}
\numberwithin{figure}{section}
\theoremstyle{plain}
\newtheorem{theorem}{Theorem}

\newtheorem{remark}{Remark}

\makeatother

\begin{document}

\maketitle

\vspace*{-2ex}
\begin{center}\small
Faculty of Mechanics and Mathematics, Moscow State University,\\ 
 Leninskie Gory 1, GSP-1,
 Moscow, 119991, Russia \\e-mail: malyshev2@yahoo.com
 \end{center}
 \vspace*{1ex}
 
\begin{abstract}
We consider stochastic models of chemical reaction networks with time
dependent input rates and several types of molecules. We prove that,
in despite of strong time dependence of input rates, there is a kind
of homeostasis phenomenon: far away from input nodes the mean numbers
of molecules of each type become approximately constant (do not depend
on time).

Key words and phrases: stochastic models of biological systems, almost
periodic input rates, stochastic networks, thermodynamic limit 
\end{abstract}

\section{Introduction and the Model}

The problem has arised from thinking over homeostasis phenomenon.
Many biological phenomena can be described in terms of the concentrations
or the mean numbers $m_{\alpha,v}$ of molecules of type $v$ in different
parts (compartments) $\alpha$ of the biological organism. Compartments
can be imagined as space separated parts of the cell or biological
organism, where some system of chemical reactions takes place. As
the result of chemical reactions in a given compartment, molecules
can change their types and leave the compartment to arrive at another
one. One of many features of the homeostasis is that large fluctuations
of external media become smoothed inside of the organism 

Consider a directed graph $G$ with $M=N_{1}J$ vertices, each vertex
is a pair $i=(\alpha,v)$, where $v=1,...,J$ represent molecule types,
and $\alpha=1,...,N_{1}$ are the compartments. We will define a continuous
time Markov or, more generally, non Markov random process, the states
of which are the vectors
 $(\xi_{\alpha,v},\alpha=1,...,N_{1};v=1,...,J)$, $\xi_{\alpha,v}\in Z_{+}$.
This process models chemical reaction networks, where the number $\xi_{\alpha,v}$
is interpreted as the number of molecules of type $v$ in the compartment
$\alpha$. Edges $((\alpha,v),(\alpha,v^{\prime}))$ of the network
correspond to some mechanism of the chemical reaction in the compartment
$\alpha$, edges $((\alpha,v),(\alpha^{\prime},v))$ correspond to
molecule transportation from the compartment $\alpha$ to the compartment
$\alpha^{\prime}$. 

This random process is defined by the following transition rates: 

\begin{enumerate}
\item Molecules of type $v$ arrive from exterior with (time dependent)
input rates $\lambda_{i}=\lambda_{\alpha,v}(t)$ to the compartment
$\alpha$. If $\lambda_{i}$ is not identically zero then the vertex
is called input vertex. Input rates are assumed to be almost periodic
(in particular, periodic) or stationary stochastic; 
\item In Markov case any molecule at the vertex $i$ with rate $\lambda_{ij}=\lambda_{\left(\alpha,v\right)(\beta,w)},$
jumps to the vertex $j$, and with rate $\lambda_{i,0}$ leaves the
network; 
\item In non Markov case another language is more convenient. After a molecule
appears at vertex $i$ , it decides immediately with (time independent)
probabilities $p_{ij}=p_{\left(\alpha,v\right)(\beta,w)},$ to go
from the compartment $\alpha$ to the compartment $\beta$, simultaneously
changing its type $v$ to $w$. Another possibility for this molecule
is to leave the system with the probability $p_{i0}$. Obviously it
should be\[
\sum_{j}p_{ij}+p_{i0}=1\]
 We assume that $G$ has (one) directed edge from $i$ to $j$ iff
$p_{ij}$ is not identically zero. After this decision, some random
time $\tau_{ij}$ is necessary until the molecule from the node $i$
appears in the node $j$ after this time. All these random times are
mutually independent and have distribution functions $F_{i,j}$, which
depend only on $i,j$. Note that in the Markov case $F_{i,j}$ are
exponential with parameter $\lambda_{ij}$. 
\end{enumerate}
Now we explain our main result intuitively together with its biological
interpretation. Our goal is to discuss homeostasis phenomenon in such
networks. General mathematical notion of the homeostasis, as one of
the most fundamental biological concepts, may not exist at all. However,
inside formal models one may suggest definitions of the homeostasis,
which could be reasonable for a given class of models. We think moreover
that the notion of the homeostasis we suggest here could concern other
applied networks. 

In most of the paper, all rates, except input rates, are assumed to
be constant (time independent). There are many reasons for such assumption.
For example, the rates of unary reactions normally depend on the concentrations
of the enzymes, catalyzing the corresponding unimolecular reaction.
Enzyme concentrations are governed by a process of enzyme synthesis
(DNA-RNA-ribosome-protein) which has slower time scale. That is why
on the properly chosen time scale these rates are approximately constant.
Moreover, in the vicinity of thermodynamic equilibrium (minimal Gibbs
free energy), the reaction rates weakly depend on time, see \cite{Mal3}. 

Roughly speaking, we prove that (in despite of strong dependence of
input rates, and thus of the concentrations in the input nodes, on
time) there is a kind of homeostasis phenomenon: far away from input
nodes the means \[
m_{i}(t)=E\xi_{\alpha,v}(t)\]
 become approximately constant (do not depend on time). In real biology
there exist, surely, more complicated control mechanisms. The goal
of this paper is to show that even such natural rough mechanism like
randomness sometimes effectively controls homeostasis.

\section{Almost Periodic Input}

A path of length $n$ from vertex $i$ to vertex $j$ is a sequence
$e_{1}=(i_{1},j_{1}),...,e_{n}=(i_{n},j_{n})$ of directed edges such
that $i_{1}=i,j_{n}=j$ and $j_{k}=i_{k+1}$for any $k=1,...,n-1$.
Denote $\mathbf{L}_{n}(i,j)$ the set of all paths from $i$ to $j$
of length $n$. 

We introduce a non negative measure $P_{n}$ on the set $\mathbf{L}_{n}(i,j)$.
For any path $L=((i_{1},i_{2}),...,(i_{n-1},i_{n})),i_{1}=i,i_{n}=j,$
this measure is \begin{equation}
P_{n}(L)=p_{i_{1}i_{2}}...p_{i_{n-1}i_{n}}\label{path}\end{equation}
 Then\[
p^{(n)}(i,j)=\sum_{L\in\mathbf{L}_{n}(i,j)}P_{n}(L)\]
 is the $n$-step transition probability for the discrete time (improper)
Markov chain on the set of vertices of the graph $G$ with the transition
probability matrix $(p_{ij})$. 

It follows from the definitions, that each molecule, independently
of others, performs random walk $\eta(t)$ on the graph, where $\eta(t)$
is a semi-Markov process. The distribution of this process is uniquely
defined by transition probabilities $p_{ij}$ and distribution functions
$F_{ij}.$ Let\[
P_{ij}(t)=P\left(\eta(t)=j|\eta(0)=i\right)\]
Here we suppose that at the initial moment the sojourn time at state
$i$ is equal to $0$ for all states $i$. Then,\begin{equation}
P_{ij}(t)=\sum_{n=1}^{\infty}\sum_{L\in\mathbf{L}_{n}(i,j)}P_{n}(L)\int_{0}^{s}dF_{L}(u)(1-F_{j}(t-u)),\label{trprob}\end{equation}
 where $F_{L}$ is the convolution of distributions $F_{i_{1},i_{2}},...,F_{i_{n-1},i_{n}}$
and\[
F_{i}(s)=\sum_{j}F_{ij}(s)p_{ij}\]

We will need some technical assumptions: 

\begin{enumerate}
\item In the first theorem we assume that $\lambda_{i}(t),-\infty<t<\infty,$
are non negative uniform almost periodic functions in the sense of
\cite{Levitan}. One of the equivalent definitions of uniform almost
periodic functions is that there exists the following limit for all
real $\sigma:$\[
c(\sigma,i)=\lim_{T\rightarrow\infty}\frac{1}{T}\int_{0}^{T}\lambda_{i}(s)\exp\left(-i\,\sigma s\right)ds\]
 and, moreover, the function $c(\sigma,i)$ is different from $0$
at most on a countable set of values $\sigma.$ Let $\sigma_{k,i}$
be the values of $\sigma,$ such that $c(\sigma_{k,i},i)\neq0.$ Put
$c_{k}(i)=c(\sigma_{k,i},i).$ Note that $c(0,i)\neq0,$ as $\lambda_{i}(s)$
is non negative, and we denote $c(0,i)=c_{0}(i)$. We will assume
that $\sum_{k=0}^{\infty}\left|c_{k}(i)\right|<\infty.$ Then the
Fourier series \begin{equation}
\lambda_{i}(t)=\sum_{k=0}^{\infty}c_{k}(i)\exp\left(i\,\sigma_{k,i}t\right),\label{fser}\end{equation}
 converges to the function $\lambda_{i}(t)$ uniformly over $-\infty<t<\infty$. 
\item There exist the means \[
\int_{0}^{\infty}sdF_{ij}(s)<\infty\]

\item Assume \begin{equation}
b(i,j)=\sum_{n>0}p^{(n)}(i,j)<\infty\label{trans}\end{equation}
For this it is necessary that the probability to leave the network
$p_{i0}>0$ at least for one $i$.
\end{enumerate}
Let $I$ be the set of input nodes. Define the following constants\[
d_{j}=\sum_{i\in I}\overline{\lambda_{i}}b(i,j)\mu_{j},\overline{\lambda_{i}}=c_{0}(i),\mu_{j}=\int_{0}^{\infty}sdF_{i}(s)\]

Introduce the class $\mathcal{M}(C,a,\delta),$ where $C,a>0$ and
$0<\delta<1$, consisting of the networks which satisfy the above
assumptions (1)-(3) and also the following conditions:

$(i)$ the following inequality holds\[
\sum_{i\in I}\sum_{k=0}^{\infty}\left|c_{k}(i)\right|\leq C;\]

$(ii)$ if $|\sigma|<a$ and $\sigma\neq0,$ then $c(\sigma,i)=0$
for all $i;$

$(iii)$ for any $a>0$ there is $0<\delta<1$ such that if $|\sigma|\geq a$
then $|\psi_{i,j}(\sigma)|\leq\delta<1,$where $\psi_{i,j}(\sigma)$
are characteristic functions of distributions $F_{ij};$

\begin{theorem}\label{t-pochti-per}Fix some network satisfying assumptions
(1)-(3). Then for any node $j$ of that network the function\begin{equation}
m_{j}^{\infty}(t)=\sum_{i\in I}\int_{0}^{\infty}\lambda_{i}(t-s)P_{ij}(s)ds\label{mean}\end{equation}
is uniform almost periodic and as $t\to\infty$\[
|m_{j}^{\infty}(t)-m_{j}(t)|\rightarrow0\]
 There exists some constant $B=B(C,a,\delta)>0$ such that for any
network from the class $\mathcal{M}(C,a,\delta)$ the inequality\[
\sup_{t}\left\vert m_{j}^{\infty}(t)-d_{j}\right\vert \leq B\delta^{N}\]
holds for any node $j$ with $dist(j,I)>N$.\end{theorem}

Proof. We use the following obvious representation, see \cite{Massey1},
for the means\[
m_{j}(t)=\sum_{i\in I}\int_{0}^{t}\lambda_{i}(s)P_{ij}(t-s)ds=\sum_{i\in I}\int_{0}^{t}\lambda_{i}(t-s)P_{ij}(s)ds\]
 By (\ref{trprob}) we have\[
\int_{0}^{\infty}P_{ij}(s)ds=\sum_{n=1}^{\infty}\sum_{L\in\mathbf{L}_{n}(i,j)}P_{n}(L)\int_{0}^{\infty}ds\int_{0}^{s}dF_{L}(u)(1-F_{j}(s-u))\]
 Integrating by parts in the above integral we get that for all paths
$L$ to the node $j$\[
\int_{0}^{\infty}ds\int_{0}^{s}dF_{L}(u)(1-F_{j}(s-u))=\int_{0}^{\infty}sdF_{j}(s)=\mu_{j}\]
 and, hence, \begin{equation}
\int_{0}^{\infty}P_{ij}(s)ds=\sum_{n=1}^{\infty}p^{(n)}(i,j)\mu_{j}=b(i,j)\mu_{j}\label{zeromem}\end{equation}
 Then, as $t\rightarrow\infty,$\[
|m_{j}^{\infty}(t)-m_{j}(t)|=\left|\sum_{i\in I}\int_{t}^{\infty}\lambda_{i}(t-s)P_{ij}(s)ds\right|\rightarrow0,\]
 as the functions $\lambda_{i}(t)$ are bounded and for each network
the set of input nodes is finite. 

Substituting Fourier series (\ref{fser}) in (\ref{mean}) and using
the uniform convergence of the Fourier series we get\begin{equation}
m_{j}^{\infty}(t)=\sum_{i\in I}c_{0}(i)\int_{0}^{\infty}P_{ij}(s)ds+\sum_{i\in I}\sum_{k=1}^{\infty}c_{k}(i)\exp\left(i\,\,\sigma_{k,i}t\right)g_{ij}(\sigma_{k,i}),\label{eqmain}\end{equation}
where \begin{equation}
g_{ij}(\sigma)=\int_{0}^{\infty}\exp\left(-i\,\sigma s\right)P_{ij}(s)ds,\label{fourtr}\end{equation}
 It follows from (\ref{zeromem}) that \begin{equation}
\sum_{i\in I}c_{0}(i)\int_{0}^{\infty}P_{ij}(s)ds=d_{j}\label{zero}\end{equation}
Let $dist(j,I)>N.$ According to (\ref{trprob}), we have \[
P_{ij}(s)=\sum_{n>N}\sum_{L\in\mathbf{L}_{n}(i,j)}P_{n}(L)G_{L}(s),\]
 where we have denoted\[
G_{L}(s)=\int_{0}^{s}dF_{L}(u)(1-F_{j}(s-u)),\]
Integrating by parts in (\ref{fourtr}) we find\[
g_{ij}(\sigma)=\frac{1}{i\sigma}\int_{0}^{\infty}\exp\left(-i\,\sigma s\right)dP_{ij}(s),\]
because $P_{ij}(0)=0$ and $P_{ij}(s)\to\infty$ as $s\to\infty.$
Note that

\[
dP_{ij}(s)=\sum_{n>N}\sum_{L\in\mathbf{L}_{n}(i,j)}P_{n}(L)dG_{L}(s)\]
and $dG_{L}(s)=dF_{L}(s)-\left(dF_{L}\star dF_{j}\right)(s).$ Therefore,

\[
g_{ij}(\sigma)=\sum_{n>N}\sum_{L\in\mathbf{L}_{n}(i,j)}P_{n}(L)\left(\frac{\psi_{L}\left(-\sigma\right)-\psi_{L}\left(-\sigma\right)\psi_{j}\left(-\sigma\right)}{i\sigma}\right),\]
where\begin{equation}
\psi_{L}\left(-\,\sigma\right)=\psi_{i_{1},i_{2}}\left(-\sigma\right)...\psi_{i_{n-1},i_{n}}\left(-\sigma\right),\label{pathf}\end{equation}
and is $\psi_{j}$ the characteristic function of distribution $F_{j}..$
Finally, we get\begin{equation}
g_{ij}(\sigma)=\sum_{n>N}\sum_{L\in\mathbf{L}_{n}(i,j)}P(L)\psi_{L}\left(-\sigma\right)\left(\frac{1-\psi_{j}\left(-\sigma\right)}{i\sigma}\right).\label{four}\end{equation}

Further, by \textbf{conditions $(ii)$} and $(iii)$ we have that
for all $\sigma_{k,i}$\[
\left|\psi_{L}\left(-\sigma_{k,i}\right)\right|\leq\delta^{|L|},\]
 where $|L|$ is the length of the path $L$ and, thus, for all $i,j,k$
with $dist(j,I)>N$ we have\begin{equation}
\left|g_{ij}(\sigma_{k,i})\right|\leq\frac{2}{a}\sum_{n>N}p^{(n)}(i,j)\delta^{n}\leq\frac{2}{a(1-\delta)}\delta^{N}\label{expes}\end{equation}
 Hence, using formulas (\ref{eqmain}),(\ref{zero}) and \textbf{condition
$(i)$} we have\[
\left|m_{j}^{\infty}(t)-d_{j}\right|\leq\sum_{i\in I}\sum_{k=1}^{\infty}|c_{k}(i)|\frac{2}{a(1-\delta)}\delta^{N}=B(C,a.\delta)\delta^{N}\]
 The theorem is proved. 

\begin{remark}Our model can be considered as an open queueing network
with $M$ nodes and infinite number of services available at each
node. Different questions for these networks were studied in \cite{Massey1,Massey3,Rol}.

However, we should note that the property of almost periodicity plays
a key role in theorem \ref{t-pochti-per}. It is not difficult to
give examples of non almost periodic functions $\lambda_{i}(t)$
for which the theorem does not hold.

\end{remark}

\section{Stochastic stationary input}

Here we assume that the input rates form a random stationary vector
process (random media) \[
(\lambda_{1}(t,\omega_{env}),...,\lambda_{|I|}(t,\omega_{env}))\]
 The components are assumed to be mutually independent, and almost
surely $|(\lambda_{i}(t,\omega_{env})|$ $\leq L_{i}$ for some constants
$L_{i}$. Then the mean numbers of molecules depend on $\omega_{env}$,
and we will study their typical behavior.

We will denote $E_{env}$ the mean over the random environment. Let
$\lambda_{i}=E_{env}\lambda_{i}(t)$ and $r_{i}(t)$ be the correlation
function: $r_{i}(t)=E_{env}(\lambda_{i}(t+s)-\lambda_{i})(\lambda_{i}(s)-\lambda_{i}),i=1,...,N.$
We assume that $r_{i}(t)$ is continuous at $0.$ Let $z_{i}(d\sigma)$
be the corresponding spectral measure, i.e.\[
r_{i}(t)=\int_{-\infty}^{+\infty}e^{it\sigma}z_{i}(d\sigma)\]
 We will need the following spectral representation \[
\lambda_{i}(t)=\lambda_{i}+\int_{-\infty}^{+\infty}e^{it\sigma}\varsigma_{i}(d\sigma),\]
 where $\varsigma_{i}(d\sigma)$ is the orthogonal random measure
such that $E_{env}\varsigma_{i}(A)=0,E_{env}\left|\varsigma_{i}(A)\right|^{2}=z_{i}(A)$
for any Borel subset $A\subset(-\infty.+\infty).$

Introduce the random process $l_{j}(t)=l_{j}(t,\omega_{env})$ which
is the mean number of molecules at node $j$ under a given $\omega_{env}$:\[
l_{j}(t)=\sum_{i\in I}\int_{0}^{t}\lambda_{i}(t-s)P_{ij}(s)ds,\]
 where $P_{ij}(s)$ is defined by formula (\ref{trprob}). Introduce
also the stationary random processes\begin{equation}
l_{j}^{\infty}(t)=\sum_{i\in I}\int_{0}^{\infty}\lambda_{i}(t-s)P_{ij}(s)ds\label{mea}\end{equation}

We define the class $\mathcal{M}_{s}(C,a,\delta),C,a>0,0<\delta<1$
consisting of networks satisfying the above condition $(iii)$ and
the following conditions

$(iv)$\[
\textrm{Var}\left(\sum_{i\in I}\lambda_{i}(t)\right)\leq C\]
or, equivalently, in terms of the spectral measure $\sum_{i\in I}\int_{-\infty}^{+\infty}z_{i}(d\sigma)\leq C.$

$(v)$ there exists spectral gap, i. e. for all $i$ the spectral
measure $z_{i}(A_{0})=0$ for some interval $A_{0}=(-a,a)$.

We define also another class $\mathcal{M}_{ss}(C,a,\delta),C,a>0,0<\delta<1,$
consisting of networks satisfying the above conditions $(iii)$,$(iv)$
and also the following condition

$(vi)$ for any $\epsilon>0$ there is $a>0$ (small enough) such
that \[
\sum_{i\in I}\int_{-a}^{+a}z_{i}(d\sigma)\leq\epsilon\]
and $b(i,j)$ and $\mu_{j}$ are uniformly bounded: $b(i,j)\leq b,$$\mu_{j}\leq\mu$
for some constants $b,\mu>0.$

It follows from this condition that the spectral measures $z_{i}(d\sigma)$
have no atoms at $0.$

\begin{theorem}Assume that conditions 2--3 hold. Then, for any $j$,
as $t\rightarrow\infty$

\[
l_{j}^{\infty}(t)-l_{j}(t)\rightarrow0\qquad with\: probability\:1.\]

There exists the constant $\hat{B}=\hat{B}(C,a,\delta)>0$ such that
for any network from the class $\mathcal{M}_{s}(C,a,\delta)$ the
inequality\[
E_{env}\left\vert l_{j}^{\infty}(t)-e_{j}\right\vert ^{2}\leq\hat{B}\delta^{2N},\qquad e_{j}=E_{env}l_{j}^{\infty}(t)\]
holds for any node $j$ with $dist(j,I)>N$. 

For any $\epsilon>0$ there exists $N_{0}$ large enough such that
for any network from the class $\mathcal{M}_{ss}(C,a,\delta)$ the
inequality\[
E_{env}\left\vert l_{j}^{\infty}(t)-e_{j}\right\vert ^{2}\leq\epsilon\]
holds for any node $j$ with $dist(j,I)>N.$ 

\end{theorem}

Proof. To prove the first assertion of the theorem let us note that\[
|l_{j}^{\infty}(t)-l_{j}(t)|=\left|\sum_{i\in I}\int_{t}^{\infty}\lambda_{i}(t-s)P_{ij}(s)ds\right|\leq\sum_{i\in I}L_{i}\int_{t}^{\infty}P_{ij}(s)ds\rightarrow0,\]
as the sum is finite and the integrals are convergent.

Prove the second assertion of the theorem. Using (\ref{mea}) and
(\ref{four}) we get the following spectral representation for the
process $l_{j}^{\infty}(t)$ \[
l_{j}^{\infty}(t)=E_{env}l_{j}^{\infty}(t)+\sum_{i\in I}\int_{-\infty}^{+\infty}e^{it\sigma}g_{ij}(\sigma)\varsigma_{i}(d\sigma),\]
 where \[
E_{env}l_{j}^{\infty}(t)=\sum_{i\in I}\lambda_{i}\int_{0}^{\infty}P_{ij}(s)ds=\sum_{i\in I}\lambda_{i}b(i,j)\mu_{j}\]
and $g_{ij}(\sigma)$ is defined by formula (\ref{fourtr}). Hence,
$\sum_{i\in I}g_{ij}^{2}(\sigma)z_{i}(d\sigma)$ is the spectral measure
of $l_{j}^{\infty}(t),$ and, thus, \[
D_{j}=E_{env}\left|l_{j}^{\infty}(t)-E_{env}l_{j}^{\infty}(t)\right|^{2}=\sum_{i\in I}\int_{-\infty}^{+\infty}g_{ij}^{2}(\sigma)z_{i}(d\sigma),\]
Let $dist(j,I)>N.$ By formula (\ref{expes}) and condition $(iv)$
we have \begin{equation}
\sum_{i\in I}\int_{|\sigma|>a}g_{ij}^{2}(\sigma)z_{i}(d\sigma)\leq\hat{B}(C,a,\delta)\delta^{2N}\label{eq:3.2}\end{equation}
 By condition $(v)$ we get\[
D_{j}=E_{env}\left|l_{j}^{\infty}(t)-E_{env}l_{j}^{\infty}(t)\right|^{2}\leq\hat{B}(C,a,\delta)\delta^{2N}\]

To prove the third assertion note that for $\epsilon>0$ arbitrary
small we can choose $\kappa>0$ small enough such that \[
\sum_{i\in I}\int_{-\kappa}^{+\kappa}g_{ij}^{2}(\sigma)z_{i}(d\sigma)\leq\epsilon\]
This follows from the condition $(vi).$ If $dist(j,I)>N_{0},$ then,
as follows from reasoning similar to (\ref{eq:3.2}), the expression
$\sum_{i\in I}\int_{|\sigma|>\kappa}g_{ij}^{2}(\sigma)z_{i}(d\sigma)$
can be made arbitrary small for $N_{0}$ large enough.

The theorem is proved.

\section{Two thermodynamic limits}

There are two quite different thermodynamic limits in our context.
The first one is quite common: instead of considering finite networks
we could consider infinite networks.

The second one correspodnds to the fact that in chemical kinetics
it is normally assumed that the number of molecules in a given compartment
is large. More exactly, we consider a family $\xi^{(\Lambda)}$ of
the processes defined above with the same graph $G$, parametrized
by the large parameter $\Lambda$ interpreted as the volume of the
compartment. For simplicity of notation we assume that all compartments
have the same volume. 

This parameter appears in our model in two places. Firstly, we assume
that the initial conditions for the means $m_{\alpha,v}^{(\Lambda)}(0)$
are such that at time zero there is a limit\[
\lim_{\Lambda\rightarrow\infty}\Lambda^{-1}m_{\alpha,v}^{(\Lambda)}(0)=c_{\alpha,v}(0)\]
 (initial concentrations). Secondly, the input rates for the process
$\xi^{(\Lambda)}$ are scaled as \[
\lambda_{i}(t)=a_{i}(t)\Lambda\]
 for some non negative functions $a_{i}(t)$, 

As the means satisfy the following equations (for the Markov case)\[
\frac{dm_{i}(t)}{dt}=\lambda_{i}(t)+\sum_{j}(m_{j}(t)\lambda_{ji}-m_{i}(t)\lambda_{ij})-m_{i}(t)\lambda_{i0}\]
 the limiting concentrations, due to linearity, \[
c_{i}(t)=\lim\Lambda^{-1}m_{\alpha,v}^{(\Lambda)}(t),\quad i=(\alpha,v)\]
exist and satisfy the same equations, see below. Note that these equations
coincide with the classical chemical kinetics equations for the system
with uni modular reactions. 

One can take into account the possibility that if a molecule goes
from compartment $\alpha$ to compartment $\beta$, it may be absent
for some time from both compartments. This case can be reduced to
the previous case if we introduce a new vertex $v_{ij}$ for any edge
$(i,j)$ with nonzero $\lambda_{ij}$. 

Instead of considering the case with general $F_{ij}$, it is more
reasonable to modify the Markov model slightly. First of all we assume
that when a particle chooses transition to a different compartment
it does not change type, that is transitions $(\alpha,v)\to(\beta,w),$
$\alpha\neq\beta,$ are only possible when $v=w$. Moreover, it appears
in compartment $\beta$ only after random time $\tau_{ik},\, i=(\alpha,v),\, k=(\beta,v)$.
All times $\tau_{ik}$ are independent and have distribution functions
$F_{ik}$. Then the limiting equations are \begin{eqnarray*}
\frac{dc_{k}(t)}{dt} & = & l_{k}(t)+\sum_{\scriptsize\begin{array}{c}
i=(\alpha,v):\\
\alpha=\beta\end{array}}c_{i}(t)\lambda_{ik}+\sum_{\scriptsize\begin{array}{c}
i=(\alpha,v):\\
\alpha\not=\beta\end{array}}\int_{0}^{t}c_{i}(t-u)\lambda_{ik}dF_{ik}(u)-\,\\
 &  & \null-\sum_{j}c_{k}(t)\lambda_{kj}-c_{k}(t)\lambda_{k0},\qquad k=(\beta,w).\end{eqnarray*}

For these cumbersome chemical kinetics equations, we can prove the
following result. 

\begin{cor}
Under conditions of theorem \ref{t-pochti-per} there exist almost
periodic functions $c_{j}^{\infty}(t),$ $j=1,\dots,J$, and constants
$d_{j}$, such that as $t\to\infty$\[
\left|c_{j}(t)-c_{j}^{\infty}(t)\right|\to0\quad and\quad\sup\left|c_{j}^{\infty}(t)-d_{j}\right|\leq B\delta^{N}\]
for $j$ satisfying the condition $dist(j,N)>N.$
\end{cor}
\begin{remark}Chemical networks were considered in many papers, some
models close to ours one can find in \cite{GadLeeOth,Mavr,Str,Tho}.
A microscopic model of one thermodynamic compartment was considered
in \cite{Mal3}. Network of compartments were introduced in \cite{Mal4}.
It is important to extend our results to more general networks. Using
deep results from \cite{RybShlos1,RybShlosVlad} one could possibly
consider networks with FIFO queues. Most important however is to study
binary reactions case.

\end{remark}

\end{document}